\documentclass[preprint,11pt]{JHEP3} 

\JHEPspecialurl{http://jhep.sissa.it/JOURNAL/JHEP3.tar.gz}
\JHEPspecialurl{http://jhep.sissa.it/JOURNAL/JHEP3.tar.gz}
\usepackage{epsfig,multicol,amsmath,slashed}

\def\beq{\begin{equation}}
\def\eeq{\end{equation}}
\def\bea{\begin{eqnarray}}
\def\eea{\end{eqnarray}}

\def\eq#1{{Eq.~(\ref{#1})}}
\def\fig#1{{Fig.~\ref{#1}}}

\newcommand{\as}{\alpha_S}

\def\vev#1{{\langle|#1|\rangle}}

\newcommand{\Lb}{\left(}
\newcommand{\Rb}{\right)}

\newcommand{\p}{I\!\!P}

\newcommand{\nb}{\mbox{nb}}
\newcommand{\fb}{\mbox{fb}}
\newcommand{\pb}{\mbox{pb}}
\newcommand{\mb}{\mbox{mb}}
\newcommand{\fm}{\mbox{fm}}

\parskip=\itemsep               

\newcommand{\h}{\frac{1}{2}}

\newcommand{\aem}{\alpha_{\footnotesize{\mbox{em}}}}
\title{\LARGE \bf   Central  exclusive\, diffractive  \, Higgs boson \, production  in\\ hadron-nucleus  and nucleus-nucleus collisions at the LHC}
\author{\large  E. ~Levin\thanks{Email: leving@post.tau.ac.il,
levin@mail.desy.de;} \,\,\,\,and \,\,J.~Miller\thanks{Email:
jeremymiller@london.com,jeremymi@post.tau.ac.il;}\,\, \\
Department of Particle Physics, School of Physics and Astronomy\\
Raymond and Beverly Sackler
 Faculty
of Exact Science\\  Tel Aviv University, Tel Aviv, 69978, Israel}

\abstract{In this paper, it is shown that in hadron-nucleus and
nucleus-nucleus collisions, the main source for central exclusive
diffractive   Higgs production is photon-photon fusion. At the LHC
energy, the total cross section for this process is about 0.6 $\pb$
(for proton-gold scattering), and 3.9\,$ \nb $ (for gold-gold
collision) while the gluon-gluon fusion leads to the value of the cross section for CED Higgs production which is about 0.1\,$\nb$ and 3.9\,$pb$ respectively.}


\preprint{  TAUP -28681/08\\
\today}
\begin{document}

\voffset0.5cm

The most important  discovery, that everybody expects at the LHC, is
the discovery of the Higgs boson.  The exclusive process which has
the best experimental signature,  as far as we know, is the  central
exclusive diffractive  (CED) Higgs production.  Having two large
rapidity gaps between the Higgs and the recoiled particles (nuclei),
this process has the minimal background from the QCD  processes
without the Higgs. However, the total cross section for CED Higgs
production turns out to be very small for proton -proton
interactions: about 3 $\fb$ for gluon-gluon fusion (see detailed
estimates by the Durham group in Ref. \cite{DG}) and about 0.1\,$
\fb$
for $\gamma \gamma$ fusion, (see Ref.\cite{JER1}).\\

In this letter we consider CED Higgs production in collisions with
nuclei, and we will show that the gluon-gluon fusion even in
proton-nucleus collisions leads to a very small cross section, while
$\gamma \gamma$ fusion gives a valuable cross section (about 0.64 pb
for proton-gold collision at the LHC).  This conclusion stems from a
striking difference in the value of the survival probabilities for
these two processes, which is negligibly small for gluon-gluon
fusion, and  of the order of unity for $
\gamma \gamma \to$ Higgs process.\\

The CED Higgs production is a process with two large rapidity gaps
(LRG), between the recoiled particles and the Higgs boson. Namely,
this reaction for proton -nucleus ( p A ) collisions looks as
follows \beq \label{RE} p\,\,\,+\,\,A\,\,\to\,\,\,p
\,\,+\,\,\mbox{$\left[LRG
\right]$}\,\,+\,\,\mbox{Higgs}\,\,\,+\,\,\,\mbox{$\left[LRG
\right]$}\,\,+\,\,A \eeq Since the mass of the Higgs boson ($M_H$)
is expected to be large (say 100 GeV or more), the typical distances
in a one parton shower interaction is short ($r \propto 1/M_H\,\,
\ll\,\,\Lambda_{QCD}$), and perturbative QCD can be used to
calculate the amplitude for the reaction of \eq{RE} (see
Refs.\cite{DG,JM}). However, the processes of two partonic shower
production will ruin the two LRG signature, as it is shown in
\fig{f1}. Therefore, we need to multiply the cross section for the
processes shown in the diagrams of \fig{f1}-a and \fig{f1}-c, by the
damping factor that we call the survival probability \cite{SPG}.
Generally speaking, we do not have a scale of hardness for the two
partonic shower production, and both short and long distances can
contribute to the calculation of the survival probability.  In the
case of hadron-hadron collisions, the most difficult part of the
calculation is related to the long distance processes, for which we
do not have a theory and have to rely on the models for the value of
the survival probability (see Refs.\cite{SPLAST,RMK} and references
therein). However, we would like to mention here that the short
distance processes for which we do have a theory (high density QCD),
can give a large contribution to the value of the survival
probability
 (see Refs.\cite{JM,BBK}).\\

For hadron-nucleus and nucleus-nucleus collisions, the main source
of the shadowing corrections is the Glauber rescattering off
different nucleons. The simple formula for the survival probability,
in this takes the following form. \cite{SPG}
 \beq \label{SP}
\vev{S^2}\,\,=\,\,\frac{\int\,d^2 b \,A_H(b)\,\,\exp\Lb-
\Omega(b,s)\Rb} {\int\,d^2 b \,A_H(b)} 
\eeq 

where $A_H(b)$ is the
impact parameter image of the hard amplitude shown in \fig{f1}-a and
\fig{f1}-c, and the opacity $\Omega$ for proton-nucleus scattering
is equal to 
\beq\label{OM} \Omega(s,b)\;\;\;=\;\;\;\sigma_{tot}\Lb
\mbox{proton-proton},s,b\Rb \,\,T_A\Lb b\Rb
\,\,\,\,\,\,\,\mbox{with} \,\,\,\,\,\,\,T_A\Lb
b\Rb\,\,=\,\,\int\,\,\rho_A\Lb r\Rb\,d z
\eeq 

where $\rho_A(r)$ is
the density of nucleons in a nucleus with mass number A, and $z$ is
the longitudinal coordinate in the beam direction. $\sigma_{tot}$ is
the total cross section of the proton-proton interaction at the
energy of interest. It should be mentioned that \eq{SP} is exact in the Glauber approach and, therefore , for hadron-nucleus and nucleus-nucleus interaction we do not have uncertainties in the formula for survival probability unlike  for the nucleon-nucleon collisions.

 \FIGURE[t]{
\centerline{\epsfig{file=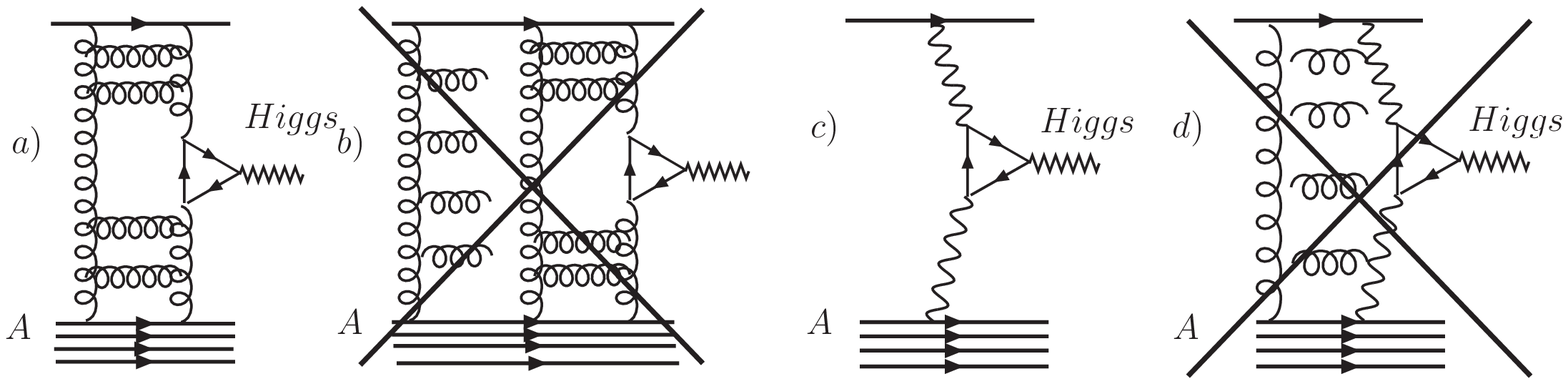,width=170mm}} \caption{ CED Higgs
production from one parton shower (\fig{f1}-a and \fig{f1}-c) and
two parton shower interaction ( \fig{f1}-b and \fig{f1}-d) that ruin
the large rapidity gap signature of the reaction of \eq{RE}.  }
\label{f1} }

 The nucleon density is given by the Wood-Saxon
parametrization (see Ref.\cite{WS}), in which it has the form
\beq \label{DENU}
\rho(b)\,\,=\,\,\frac{\rho_0}{1 - \exp\Lb \frac{r - R_A}{h}\Rb}\,\,\,\,\,\mbox{with}\,\,\,\,\,\int\,d^3\,r \rho(r)\,\,\,=\,\,\,A
\eeq
$r \,=\sqrt{b^2 + z^2}$ and $R_A \,=\, r_0\, A^{1/3}$ with $r_0\,=\,1.09\,\fm$  while $h$ does not depend on $A$. For the gold
$R_{Au}\,\,=\,\,6.38\,\fm\,$ and $h \,\,=\,\, 0.535$ (see Ref. \cite{WS}).
$\rho_0$ has a physical meaning of the nucleon density in the nuclear matter and it  can be found from the normalization, given in \eq{DENU}, which leads  to  $\rho_0 =0.17 \, \fm^{-3}$.

 Taking $\sigma_{tot}\Lb \mbox{proton-proton},s,b\Rb  = 110
\,\mb$ at the LHC energy \cite{TXS},    one can see that
$\Omega(s,b)$ in \eq{OM} is equal to
\beq \label{OM1} \Omega\Lb s_{LHC},b \Rb
\;\;\;\;=\;\;\;\;\,11\,\fm^2 \,\,T_A(b)
\eeq

 The $b$ dependence of $\exp\Lb -
\Omega\Rb$  for proton-gold collision  is shown in \fig{f3}. The striking
difference between the value of the survival probability for CED
Higgs production  for gluon-gluon fusion and $\gamma-\gamma$ fusion
stems from the quite different behavior of the hard amplitudes $A_H$
for these two processes. In the case of gluon fusion, the hard
amplitude decreases steeply with $b > R_A$,  while for photon fusion this
amplitude only slowly falls down with increasing $b$. This
qualitative difference is obvious from \fig{f3}, where the
normalized $A_H$ for both processes are plotted. The normalized
$\tilde{A}_H$ is defined as \footnote{We will use below the notation
$A_H$ instead of $\tilde{A}_H$, but we hope that it will not cause
any confusion.}

 \beq \label{NORMA}
\tilde{A}_H(b)\,\,\,\equiv\,\,\frac{ A_H(b)} {\int\,d^2 b \,A_H(b)}
\eeq

 In the case of  gluon-gluon fusion, it is better to say that for Pomeron-Pomeron  ($\p\,\p$)
fusion, the hard  amplitude takes the form (see \fig{f2}-a
)\footnote{Pomerons in \fig{f2}-a correspond to the gluon ladders.}

\beq \label{AH1p} A^{\p\p}_H(Q)\,\,\,=\,\,\int\,d^2 q_{1,
\perp}\,\int\,d^2 q_{2, \perp}\,M\Lb \p\p \to
H;\vec{q}_1,\vec{q}_2\Rb\, \,M^*\Lb \p\p \to H;\vec{q}_1 +
\vec{Q},\vec{q}_2 - \vec{Q}\Rb \eeq

where the amplitude  $M\Lb \p\p \to H;\vec{q}_1,\vec{q}_2\Rb\, $ for
CED Higgs production through $\p\p$ fusion takes the form
\cite{DG,JM}

\bea &&M\Lb \p\p \to
H;\vec{q}_1,\vec{q}_2\Rb\,=\,\frac{2}{9}\,A\,s\,G_p (q^2_1)\,G_A(q^2_2)\,\,\int\!\frac{d^{2} q_{\bot}}{q^{2}_{\bot}}\frac{(\vec{q}_{1,\bot} - \vec{q}_{\bot}) \cdot (\vec{q}_{2,\bot} + \vec{q}_{\bot})}{(\vec{q}_{1,\bot} - \vec{q}_{\bot})^2 \, (\vec{q}_{2,\bot} + \vec{q}_{\bot})^2}\,\,8\alpha{}_{s}^{2}\left(q^{2}\right)\notag \\
&&\mbox{where}\,\,\,\,\,
A\,=\,\frac{2}{3}\frac{\as\,G_F^{1/2}\,2^{1/4}\,}{\pi}\label{E:2.1}
\eea

In \eq{E:2.1}, one takes into account the proton couplings to the
gluon ladder (see \fig{f2}), by including the two gluon form factors
($G_p(q^2_1)$ and $G_A(q^2_2)$) for the gluon density in the proton
and the nucleus respectively. As it was shown in Ref. \cite{DG}, the
momentum $\vec{q}_{\bot}$ of the $t$-channel gluon in \fig{f2}-a
turns out to be large ($ q_{\bot} \ll 1/R_p$
 where $R_p$ is the radius of the proton.  In the Glauber approach,
  we consider $R_A \,\,\gg\,R_p \gg 1/q$, and therefore,
  the $Q$ dependence of
\eq{AH1p} is determined by \beq \label{APQ} \int d^2\,q_{2
\bot}\,G_A\Lb q^2_{2 \bot}\Rb\,G_A\Lb ( \vec{q}_{2 \bot} +
\vec{Q})^2 \Rb\,\,\,=\,\,\int \,d^2 b\,\,e^{ i \vec{Q} \cdot
\vec{b}}\,\, T^2_A(b) \eeq

\eq{APQ} means that  the amplitude $A_H(b)$ in \eq{SP} is
proportional to $T^2_A(b)$, and the survival probability is equal to
\beq \label{SPGG} \vev{S^2}_{GG \to H}\,\,\,\,=\,\,\,\frac{\int d^2
b\,\,T^2_A(b) \,\,\exp\Lb -\Omega(b) \Rb}{ \int d^2 b\,T^2_A(b)}
\eeq \fig{f3} shows why the value of $\vev{S^2}_{GG \to H}$ should
be small, since only in the vicinity of $b \to R_A$ does the
numerator contribute significantly in \eq{SPGG}. The calculation
gives $\vev{S^2}_{GG \to H}\,\,=\,\,8\times\, \,10^{-4}$ for proton
-gold collisions at the LHC energy.



The situation with the hard amplitude for photon fusion is quite the
opposite, namely it is a smooth function of $b$, which only slowly
decreases at large values of $b$. Therefore, the ratio of \eq{SP}
can be evaluated in the following way for CED Higgs production
through $\gamma - \gamma$ fusion \beq \label{SPG} \vev{S^2}_{\gamma
\gamma \to H}\,\,=\,\,\frac{\int\,d^2 b \,A_H(b)\,\,\exp\Lb-
\Omega(b,s)\Rb} {\int\,d^2 b
\,A_H(b)}\,\,\approx\,\,\frac{\int^{\infty}_{R^2_A}\,d b^2 \,A_H(b)}
{\int\,d^2 b \,A_H(b)}
\eeq

 \FIGURE[t]{
\centerline{\epsfig{file=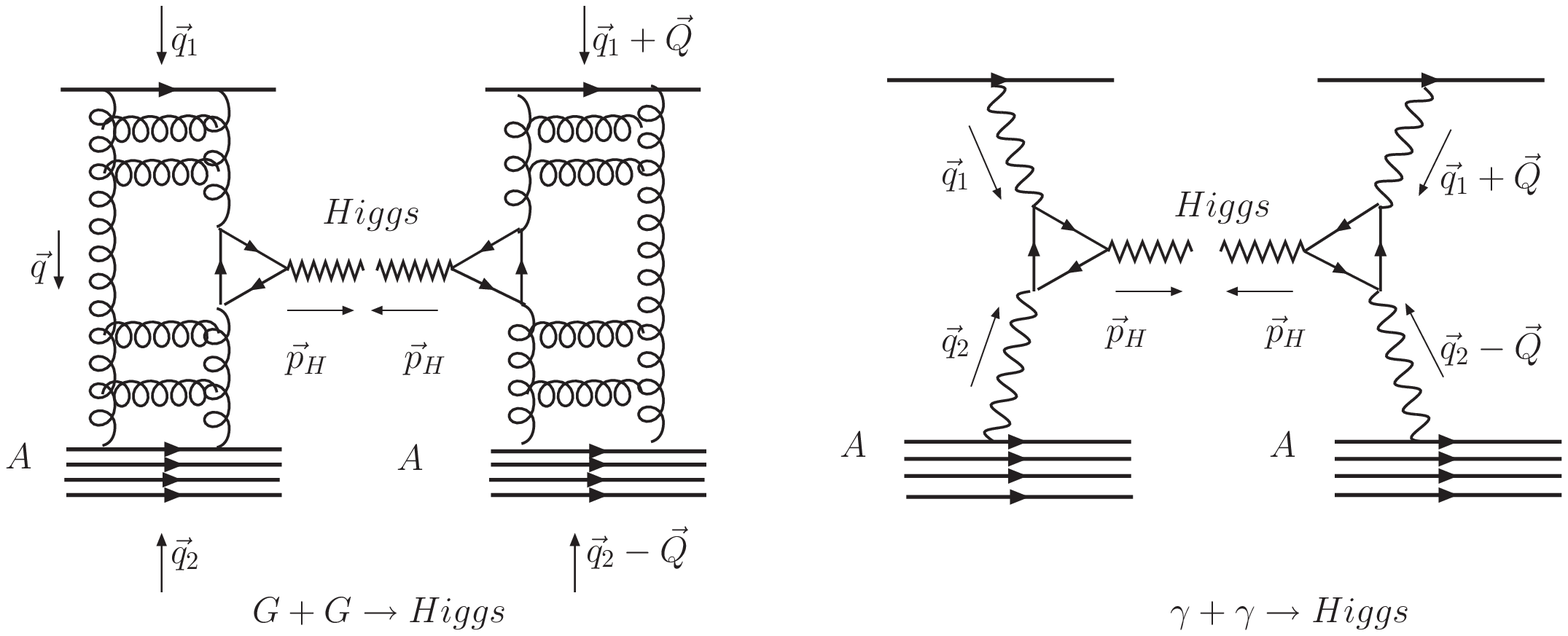,width=170mm}} \caption{The
amplitude $A_H(Q)$ for CED Higgs production due to gluon-gluon and
$\gamma-\gamma$ fusion.}  \label{f2} }

 It is clear from the
discussion above that we need to know the dependence of the hard
amplitude on $b$ at large $b$, which is intimately related to the
behavior of this amplitude at small values of $Q$ (see \fig{f3}).
The required hard amplitude can be written in the form \beq
\label{AH1} A_H(Q)\,\,\,=\,\,\int\,d^2 q_{1, \perp}\,\int\,d^2 q_{2,
\perp}\,M\Lb \gamma\gamma \to H;\vec{q}_1,\vec{q}_2\Rb\, \,M^*\Lb
\gamma\gamma \to H;\vec{q}_1 + \vec{Q},\vec{q}_2 - \vec{Q}\Rb \eeq
where (see Ref.\cite{JER1} and references therein): \beq \label{AH2}
M\Lb \gamma\gamma \to H;\vec{q}_1,\vec{q}_2\Rb\,\,\,=\,\,\,\frac{4
\pi \aem}{ q^2_{1, \perp}\, q^2_{2, \perp}}\,\, F_A\Lb
q^2_{2,\perp}\Rb\,\,F_p \Lb q^2_{1,\perp}\Rb \,\, \frac{2
s}{M^2_H}\,4\,q^{\mu}_{1, \perp}\, q^{\nu}_{2, \perp}\,\,A_{\mu \nu}
\eeq
 with \beq \label{AH 2}
 A^{\mu \nu}\,\,\,=\,\,\,\frac{8}{27}\,\frac{\aem\,G^{\h}_{F}\,2^{\frac{1}{4}}}{\pi}\,\,\left\{ q^{\mu}_1\,q^{\nu}_2\,\,\,-\,\,\,\frac{M^2_H}{2}\,g^{\mu \nu}\right\}\,\,\,\equiv\,\,A\,\,\left\{ q^{\mu}_1\,q^{\nu}_2\,\,\,-\,\,\,\frac{M^2_H}{2}\,g^{\mu \nu}\right\}
\eeq
 In \eq{AH2} $F_A(q^2_{2,\perp})$ and $F_p \Lb
q^2_{1,\perp}\Rb $ stand for the electromagnetic form factors of the
nucleus and the proton, respectively. In the region of small $Q$,
(namely  $Q \,< \,q_1\,\,\, \mbox{and}\,\,\, \,q_2$), \eq{AH1} can
be re-written as

\beq \label{AH4} A_H(Q)\,\,\,=\,\,4 \pi^2\, s^2 \,\Lb 4 \pi
\aem\,\Rb^2\,A^2\,\int^{4/R^2_p}_{Q^2}\,\frac{dq^2_{1,
\perp}}{q^2_{1, \perp}}\,\,\int^{4/R^2_A}_{Q^2}\frac{d q^2_{2,
\perp}}{q^2_{2, \perp}}\,\,\,=\,\,C\ln\Lb Q^2R^2_A/4\Rb\,\,\ln\Lb
Q^2R^2_p/4\Rb\,\,
 \eeq
where the coefficient $C$ is defined such that $C$ absorbs all
factors, since they do not contribute in the calculation of the
survival probability (see \eq{SP}). The hard amplitude in impact
parameter space is equal to \bea \label{AH5}
A_H\Lb \gamma \gamma \to H;b \Rb\,\,&=&\,\,\frac{1}{(2 \pi)^2}\,\int\,d^2 Q\,\,e^{-i\vec{b}\cdot\vec{Q}}\,\,A_H(Q)\,\,\\
&=&\,\,\,\frac{C}{(2 \pi)^2}\,2\pi
\int\,d Q^2\,J_0(b Q)\,\ln\Lb Q^2R^2_A/4\Rb\,\ln\Lb Q^2R^2_p/4\Rb\,\,\notag \\
&\xrightarrow{ 1/Q_{min} > b > R_A/2}&  \,\,\,\frac{C}{(2
\pi)^2}\,\pi\,\,\left\{\,2\,\frac{\ln\Lb
4\,b^2/R^2_A\Rb}{b^2}\,\,+\,\,\ln\Lb R^2_A/R^2_p \Rb \,\right\}
\notag \eea

where $Q_{min}= m M_H/\sqrt{s}$ (see  for example Ref.\cite{JER1}).
In the region of small $b$, we need to derive the exact expression
of \eq{AH2}, and the behavior of the hard amplitude versus $b$ is
shown
in \fig{f3}.\\

Using \eq{SPG}, we can estimate the value of the survival
probability which turns out to be equal to 0.8 $\div $ 0.85   in
proton-gold
collisions.\\

 \FIGURE[t]{
\centerline{\epsfig{file=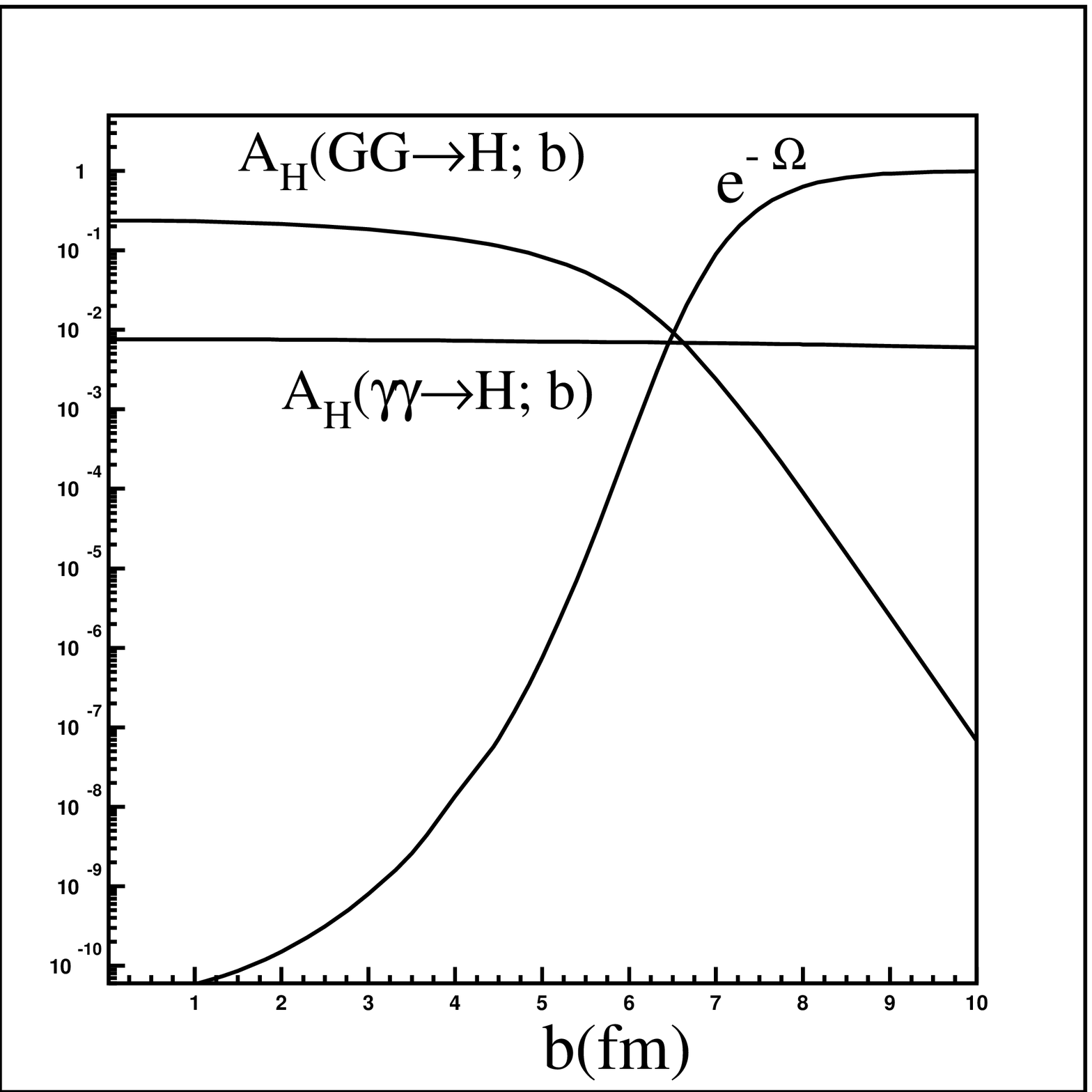,width=120mm}} \caption{The damping
factor $\exp\Lb - \Omega \Rb$,  and hard amplitude $A_h(b)$ for
$gluon\,\,\, +\,\,\, gluon\,\,\to\,\,\,Higgs$  and  $\gamma\,\,\,
+\,\,\,\gamma\,\, \to\,\,Higgs  $ fusions versus impact parameter
$b$ in the model of \eq{OM1} for the density of nucleons in a
nucleus.  }  \label{f3} }

From our estimates for the survival probability, we can obtain the
value of the cross section for CED Higgs production. Indeed, \bea
 \sigma_{pA}\Lb  G + G  \to H; s, Y_H \Rb\,&=&\,\,\sigma_{pp}\Lb  G + G  \to H; s, Y_H \Rb\,\,A^2\,\,\vev{S^2\Lb G + G  \to H\Rb}\,\,\notag \\
&=& \,\,\,33 \times\,\sigma_{pp}\Lb  G + G  \to H; s, Y_H \Rb\,\,;\label{XS1}\\
\notag\\
\sigma_{pA}\Lb  \gamma + \gamma   \to H; s, Y_H \Rb\,&=&\,\, \sigma_{pp}\Lb  \gamma + \gamma  \to H; s, Y_H \Rb\,\,Q_A^2\,\vev{S^2\Lb \gamma + \gamma \to H\Rb} \notag\\
&=&\,\,5 \times\,10^3 \,\times \, \sigma_{pp}\Lb  \gamma + \gamma
\to H; s, Y_H \Rb;\label{XS2} \eea
 
 where $Q_A$ ($A$) is the
number of protons (nucleons)  in the nucleus  and $Y_H$ is the
rapidity of the Higgs boson. The numbers in \eq{XS1} and \eq{XS2},
are given for proton-gold collisions at the LHC energy. One can see,
that the cross section for gluon fusion is extremely small, while in
the case of photon fusion, it can lead to a measurable cross
section. Taking for the value of $ \sigma_{pp}\Lb \gamma +\gamma \to
H; s, Y_H \Rb \,=\,0.12\, \fb$ (see Ref.\cite{JER1}) we obtain for $
\sigma_{pA}\Lb \gamma + \gamma \to H; s, Y_H \Rb = 0.64\,\, \pb$ at
the LHC energy. For gluon-gluon fusion we have $\sigma_{pp}\Lb  G +
G  \to H; s, Y_H \Rb\,\,=\,\,3\,$fb which leads to the value of  $
\sigma_{pA}\Lb  G + G  \to H; s, Y_H \Rb$ for proton-gold collision
at the LHC enegy about 100 \,$\fb$, which is in 6 times 
smaller than the cross section for photon fusion.
  \\

The same pattern in estimates can be seen for CED Higgs production
in ion-ion collisions.   In \eq{OM} we need to replace $T_A(b) $ by
the overlapping integral of  profile functions for two nuclei,
namely \beq \label{REPLCE} T_A(b)\,\,\longrightarrow\,\,\,T_{A_1
A_2} \,\,\,=\,\,\, \int d^2 b' T_{A_1}\Lb b'^{2^{}}
\Rb\,\,T_{A_2}\Lb (\vec{b} - \vec{b}')^2 \Rb \eeq For gluon-gluon
fusion the hard amplitude $A_H(b)$  has the form

\bea \label{GGAH} &&A_H(b)\,\,\,\propto\,\, \int d^2 b'\,
T^2_{A_1}\Lb b'^{2^{}}\Rb \,\,T^2_{A_2}\Lb (\vec{b} - \vec{b}')^2
\Rb
\,\,\,\,\mbox{which is a Fourier image of}\,\,\,\\
&&\int d^2 q_{1, \bot}\,G_{A_1}( q_{1 \bot})\,\,G_{A_1}( \vec{q}_{1
\bot}  + \vec{Q})\,\times \,\int d^2 q_{2, \bot}\,G_{A_2}( q_{2
\bot})\,\,G_{A_2}( \vec{q}_{2 \bot}  + \vec{Q}) \notag \eea

The survival probability for gluon fusion is small ( $\approx 8.16\,10^{-7}$), for gold-gold scattering). This survival probability leads to the value of the cross section in the case of gold - gold scattering 
$3 \,\fb \,\times  \,8\,10^{-7} \,\times\, A^2\,\,=\,\,3.92 \,\pb$.
However, for photon fusion the value of the survival probability
does not change too much, while the cross section is proportional to
$Q^4_A$, leading to an enhancement in the value of the cross section
for CED Higgs production in gold-gold collisions at the LHC, by a
factor of $  3.9 \,\times\,10^7$. In other words, the value for
$\sigma_{AA}\Lb \gamma + \gamma \to H; s, Y_H \Rb \,\approx 3.9
\,\nb$ for gold-gold collisions.

Therefore, we can conclude that central exclusive diffractive Higgs
production in the case of hadron-nucleus collisions, as well as in
the case of nucleus-nucleus collisions, goes through the reaction\\
$\gamma + \gamma \to $ Higgs, and can reach a rather significant
value for the LHC energies:
 $ \sigma_{AA}\Lb  \gamma + \gamma   \to H; s, Y_H \Rb\,\, \approx 3.9\,\, \nb$ for gold-gold collisions, and
  $\sigma_{pA}\Lb  \gamma + \gamma   \to H; s, Y_H \Rb \approx 0.64\,\, \pb$.
 The advantage of the photon fusion reaction, is the fact that we can calculate its cross section without any
 theoretical uncertainty. This makes the collision with nuclei
 a valuable tool for the discovery of the Higgs boson at the LHC.

\section* {Acknowledgments}
We are grateful to Errol Gotsman, Lev Lipatov, Uri Maor and Alex
Prygarin   for fruitful discussions on the subject.  This research
was supported in part by the Israel Science Foundation, founded by
the Israeli Academy of Science and Humanities, by BSF grant $\#$
20004019 and by a grant from Israel Ministry of Science, Culture and
Sport and the Foundation for Basic Research of the Russian
Federation.

\end{document}